\documentclass[sigconf]{acmart}
\AtBeginDocument{%
  }

\copyrightyear{2026}
\acmYear{2026}
\setcopyright{cc}
\setcctype{by}
\acmConference[ICSE-FoSE]{2026 IEEE/ACM 48th International Conference on Software Engineering: Future of Software Engineering }{April 12--18, 2026}{Rio de Janeiro, Brazil}
\acmBooktitle{2026 IEEE/ACM 48th International Conference on Software Engineering: Future of Software Engineering (ICSE-FoSE), April 12--18, 2026, Rio de Janeiro, Brazil}

\begin{document}

\title{Towards A Sustainable Future for Peer Review in Software Engineering}

%%
%% The "author" command and its associated commands are used to define
%% the authors and their affiliations.
%% Of note is the shared affiliation of the first two authors, and the
%% "authornote" and "authornotemark" commands
%% used to denote shared contribution to the research.

\author{Esteban Parra}
\orcid{0000-0001-9813-9518}
\email{esteban.parrarodriguez@belmont.edu}
\affiliation{%
  \institution{Belmont University}
  \city{Nashville}
  \state{Tennessee}
  \country{USA}
}

\author{Sonia Haiduc}
\orcid{0000-0001-8793-8293}
\email{shaiduc@fsu.edu}
\affiliation{%
  \institution{Florida State University}
  \city{Tallahassee}
  \state{Florida}
  \country{USA}
}

\author{Preetha Chatterjee}
\orcid{0000-0003-3057-7807}
\email{preetha.chatterjee@drexel.edu}
\affiliation{%
  \institution{Drexel University}
  \city{Philadelphia}
  \state{Pennsylvania}
  \country{USA}
}

\author{Ramtin Ehsani}
\orcid{0000-0003-1517-7135}
\email{ramtin.ehsani@drexel.edu}
\affiliation{%
  \institution{Drexel University}
  \city{Philadelphia}
  \state{Pennsylvania}
  \country{USA}
}

\author{Polina Iaremchuk}
\orcid{0009-0009-4298-1170}
\email{polina.iaremchuk@bruins.belmont.edu}
\affiliation{%
  \institution{Belmont University}
  \city{Nashville}
  \state{Tennessee}
  \country{USA}
}

%%
%% By default, the full list of authors will be used in the page
%% headers. Often, this list is too long, and will overlap
%% other information printed in the page headers. This command allows
%% the author to define a more concise list
%% of authors' names for this purpose.
\renewcommand{\shortauthors}{Trovato et al.}

%%
%% The abstract is a short summary of the work to be presented in the
%% article.
\begin{abstract}
  Peer review is the main mechanism by which the software engineering community assesses the quality of scientific results. However, the rapid growth of paper submissions in software engineering venues has outpaced the availability of qualified reviewers, creating a growing imbalance that risks constraining and negatively impacting the long-term growth of the Software Engineering (SE) research community. 
  
  Our vision of the Future of the SE research landscape involves a more scalable, inclusive, and resilient peer review process that incorporates additional mechanisms for: 1) attracting and training newcomers to serve as high-quality reviewers, 2) incentivizing more community members to serve as peer reviewers, and 3) cautiously integrating AI tools to support a high-quality review process.
\end{abstract}

%%
%% The code below is generated by the tool at http://dl.acm.org/ccs.cfm.
%% Please copy and paste the code instead of the example below.
%%

%%
%% Keywords. The author(s) should pick words that accurately describe
%% the work being presented. Separate the keywords with commas.
%% A "teaser" image appears between the author and affiliation
%% information and the body of the document, and typically spans the
%% page.

%%
%% This command processes the author and affiliation and title
%% information and builds the first part of the formatted document.
\maketitle

\section{Introduction}

Peer review is the backbone of trust, legitimacy, and collective knowledge-building in software engineering. Similarly to other disciplines, the decision on whether to accept a paper for a journal or a conference lies with editors and Program Committee (PC) chairs, respectively \cite{Lin2023}. However, due to the volume of submissions and to avoid biases in the decision-making process, editors and chairs make their decision primarily based on the reviews provided by a set of expert reviewers \cite{Lin2023}.
%% However, due to the high volume of submisions, editors and chairs focus primarily on a set of expert reviews which aid in accepting high qualiy papers and facilitates avoiding biases in decision making.

Reviewers are trusted individuals invited to provide professional reviews for papers based on their expertise and standing in the research community. Generally speaking, reviewers are assigned a set of papers and have the following core responsibilities for each paper: reading the paper, assessing the paper's quality and fit for publication in the venue using various criteria (e.g., novelty, soundness, relevance, presentation, etc.), giving a recommendation for acceptance or non-acceptance of the work, and providing clear and constructive feedback to the authors.
%%Reviewers are faced with high professional expectations when it comes to following standarts, since they have generally proven their expertise and earned the respect in the research commuty. Once submission round has been completed, they are assigned a set of papers and have the following core responsibilities for each paper: reading the paper, assessing the paper's quality and fit for publication in the venue using various criteria (e.g., novelty, soundness, relevance, presentation, etc.), giving a recommendation for acceptance or non-acceptance of the work, and providing clear and constructive feedback to the authors.

Despite reviewers' work being essential for the software engineering research community, reviewing research papers is unpaid, volunteer work performed by members of the community in addition to their multiple tasks as researchers and/or educators \cite{Trinkenreich2025}. Submission volumes have grown faster than reviewer availability, leading to high reviewer load, with some members of the community conducting over 20 reviews per year for conferences alone \cite{Ernst2021}. Furthermore, reviewing is a time-consuming task; for example, 88\% of reviewers spend over 2 hours to read and review a single journal paper \cite{Ernst2021}. This can lead reviewers to work on the weekends or late in the evenings \cite{Ernst2021}. At the same time, reviewing for the main tracks of larger conferences can be a year-long commitment, limiting reviewer availability for other venues or tracks.

\begin{figure} [ht!]
    \centering
    \footnotesize    \includegraphics[width=\linewidth]{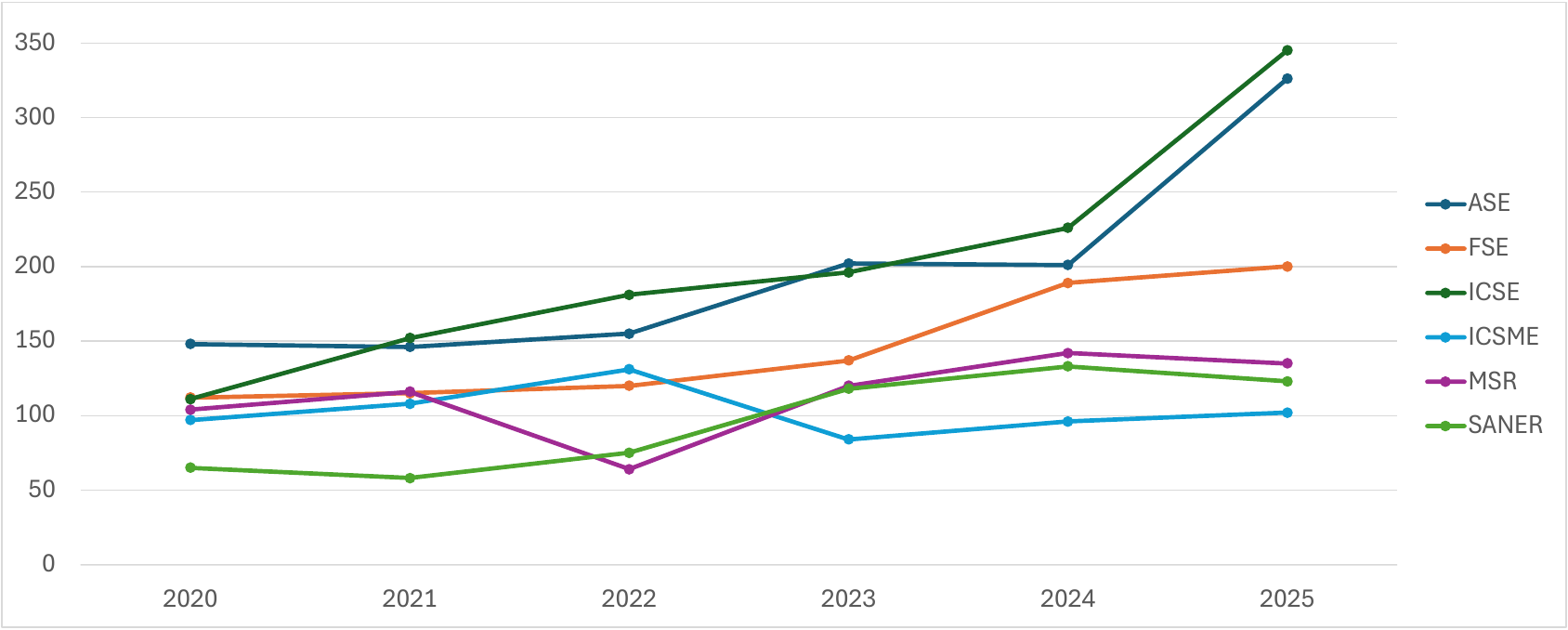}
            \captionsetup{font=small}
    \caption{PC Size Over Time}
    \label{fig:PCSize}
\end{figure}

\begin{figure} [ht!]
    \centering
    \footnotesize    \includegraphics[width=\linewidth]{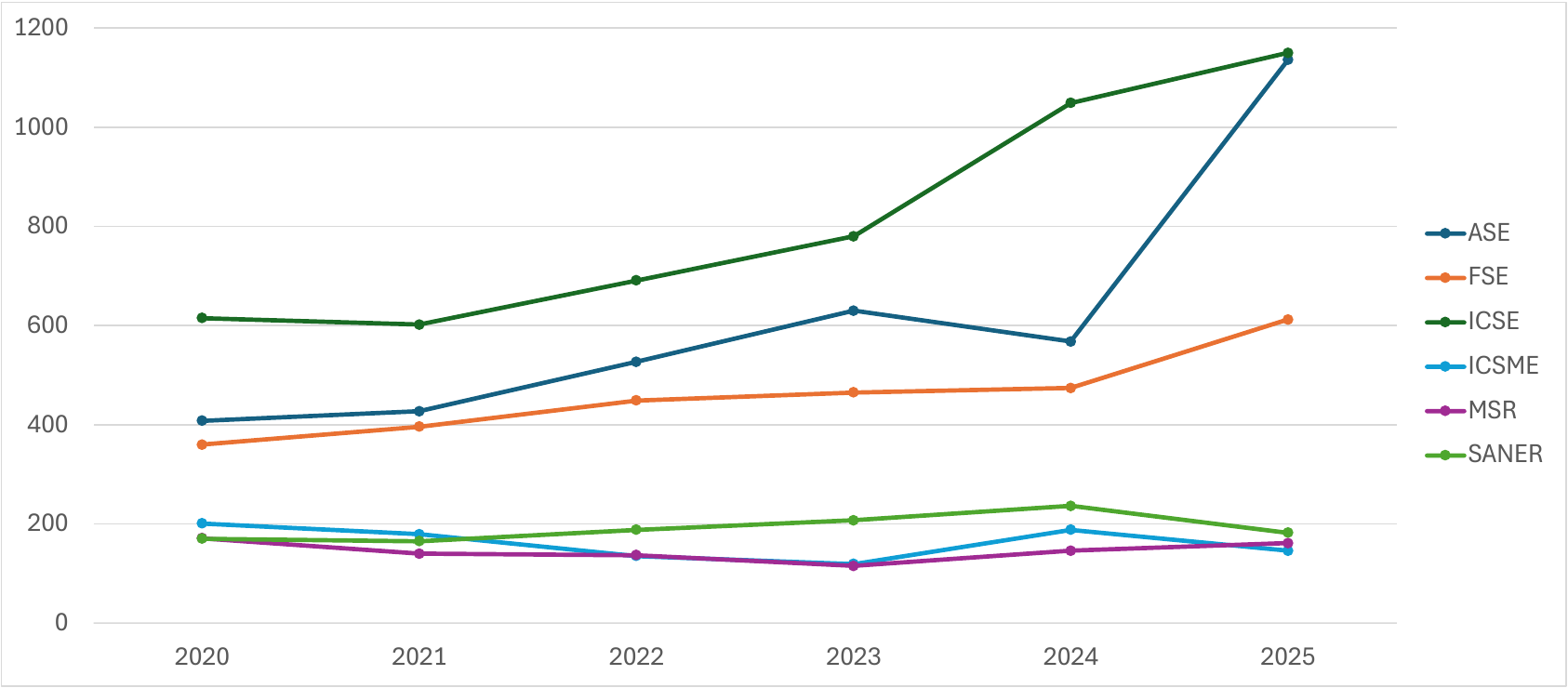}
            \captionsetup{font=small}
    \caption{Submissions Over Time}
    \label{fig:submissions}
\end{figure}

Figures \ref{fig:PCSize} and \ref{fig:submissions} show the changes in the last five years with respect to the size of the program committee and number of submissions to the main tracks of some of the largest software engineering conferences ranked A* and A (i.e., ICSE, FSE, ASE, ICSME, MSR, and SANER).
 
Figure \ref{fig:submissions} is the number of main track submissions that underwent peer review (i.e., without desk rejections). The figure shows that in the last two years, the number of submissions has increased significantly. To account for this increase in submissions, the largest conferences have also increased their PC size. However, even with the increased PC size, the reviewer workload has often increased given that each paper is usually assigned at least three reviewers. For example, for ASE, the size of the main track PC has more than doubled between 2020 and 2025 (from 148 to 326 PC members), while the number of main track submissions has nearly tripled (408 in 2020 vs 1136 in 2025), leading to a workload increase from 8.27 to 10.45 papers per PC member on average. Furthermore, many of the community members serve on the PC of multiple conferences or on the PCs of multiple tracks of the same conference, leading to an overall increasing review load.

As it stands, the discrepancy between the growth in submissions and the availability of reviewers is not sustainable and risks harming the quality of the reviewing process and the potential for growth of the SE community. We envision a future of SE research in which quality peer review is made sustainable by expanding the pool of high-quality reviewers through scalable training, retaining, and incentivizing both established community members and newcomers to serve as peer reviewers.

\section{What Is Working}

The feedback provided by reviewers on the quality of papers is incredibly valuable, as it allows authors to expand and strengthen their work before it is published. Even in cases of non-acceptance, the feedback provides actionable steps from experienced researchers on potential critical issues that, when addressed, should improve the paper for future submissions. Furthermore, the variety of tracks with different review criteria, such as the Reproducibility Studies and Negative Results at SANER\footnote{\url{https://conf.researchr.org/track/saner-2026/saner-2026-reproducibility-studies-and-negative-results-rene-track}} and ICSME\footnote{\url{https://conf.researchr.org/track/icsme-2026/icsme-2026-replication-and-negative-results}}, increases venue accessibility to a wider range of work and is a valuable tool for growing the community.

\subsection {Distinguished Reviewer Awards}

Peer review is viewed by members of the community as a shared form of quality control. Although the reasons for accepting to be a reviewer vary, many reviewers accept this role from a sense of professional duty. 

The main mechanism used by most software engineering research conferences to recognize the reviewers' labor is through the Distinguished Reviewer Award. This award is presented to a subset of reviewers who have made the greatest contributions to the review process \cite{Trinkenreich2025}. This type of award represents a valuable recognition of a researcher's efforts in the scientific community and an addition to the CV that could support their career goals \cite{Trinkenreich2025}. Therefore, it can serve as an incentive to participate in the reviewing process. However, given that very few reviewers get this recognition, it should ideally be complemented by other types of incentives for recruiting reviewers. 

\subsection {Junior/Shadow PC}

The 18th edition of the IEEE/ACM International Conference on Mining Software Repositories (MSR'21) established the Shadow PC mentorship program, later renamed Junior PC, as a mechanism to provide opportunities for early-career researchers, namely PhD students, post-docs, new faculty members, and industry practitioners, to learn about and get involved in the academic peer review process, aiming to increase the pool of qualified reviewers \cite{Thongtanunam2021}. The members of the Shadow/Junior PC are integrated in the peer review process with the members of the main technical track PC using a 2-1 model, where each paper is reviewed by 2 members of the regular PC and 1 member of the Junior PC. This way, Junior PC members have the opportunity to observe and learn from the reviews of more seasoned members of the community and can also receive feedback on their own reviews. 

The Junior PC mentorship program continues to be a part of MSR\footnote{\url{https://2026.msrconf.org/track/msr-2026-junior-pc}} to this day, as it has proven to be an effective mechanism to increase the reviewer pool in the software engineering research community. Furthermore, a similar mentorship program, the Shadow PC, has been incorporated in ICSE\footnote{\url{https://conf.researchr.org/track/icse-2026/icse-2026-shadow-research-track-program-committee}}, providing an opportunity for early-career researchers and graduate students who have not yet served on a technical research track program committee at international SE conferences to learn and contribute to the peer review process in one of the most prestigious venues in the field \cite{Varghese2024}. 

\section{What Is Not Working}
Reviewing can be invisible labor with little career recognition in a “Publish or Perish” research culture, which makes it challenging for the reviewer pool to expand organically.

\subsection {Scale and Overload}

While peer review is crucial for the community, in practice it is dependent on the community members' workload and availability. To make the process scalable and sustainable, the commonly referenced goal is "to review as much as you are reviewed" \cite{Ernst2021}. This means that for each paper an author submits, that author would be expected to perform three reviews in return (assuming an average of three reviewers per paper). However, due to the disjoint nature of the venues, there is currently no mechanism to track or enforce this practice.

\begin{figure} [ht!]
    \centering
    \footnotesize    \includegraphics[width=\linewidth]{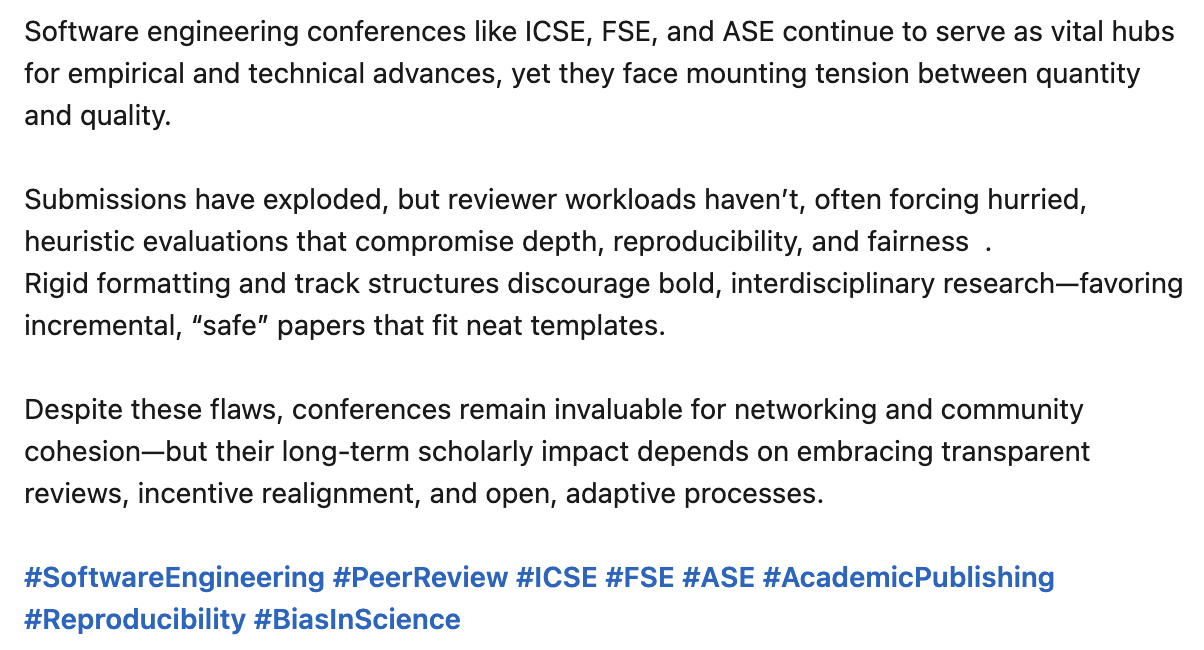}
            \captionsetup{font=small}
    \caption{A Linked-In Post on Peer Review in SE Venues}
    \label{fig:anecdote}
\end{figure}

Notably, one of the growing pains in the SE research community is that submission volumes have grown faster than reviewer capacity, as the availability of quality researchers to serve as reviewers for papers in both conferences and journals is limited, leading to frustration from reviewers and authors alike \cite{Briand2025, Tafreshipour2025}. Figure \ref{fig:anecdote} presents a recent LinkedIn post from a member of the software engineering research community that echoes this sentiment and further highlights potential concerns with respect to the quality of the reviews.

Advances in the natural language capabilities of Large Language Models (LLMs) and Generative Artificial Intelligence (Gen-AI) have enabled these tools to also generate research paper reviews \cite{Lin2023, Checco2021}. There are growing concerns regarding the practice of some reviewers to generate and submit AI reviews with little or no informed editing or corrections, particularly given the fact that AI tools lack expertise in the field and may even replicate existing biases \cite{Checco2021, Thai2025, Naddaf2025B}. In addition, providing a manuscript under review to an AI tool can lead to violating the confidentiality of the peer review process, given that AI tools can "learn" from the content of the manuscript. As a consequence of these concerns, ACM and IEEE currently have policies in place regarding the use of generative AI or LLMs in peer review. These guidelines specify that "uploading any part of a submission to an LLM or other third-party system that does not promise to maintain the confidentiality of that information is not permitted" \cite{ACMpeer, IEEEpeer}.

Despite existing guidelines limiting the use of AI in the peer review process, recent cases in computer science \cite{Naddaf2025B} and other disciplines \cite{Naddaf2025A} have brought to the forefront issues regarding the rising proliferation of low-quality reviews produced by AI in peer reviewing. Attributes of low-quality reviews resulting from reviewers' use of AI include: fake/hallucinated citations, suspiciously long and vague feedback, and methodologically incorrect suggestions that do not make sense \cite{Naddaf2025B}.

\begin{figure} [ht!]
    \centering
    \footnotesize    \includegraphics[width=\linewidth]{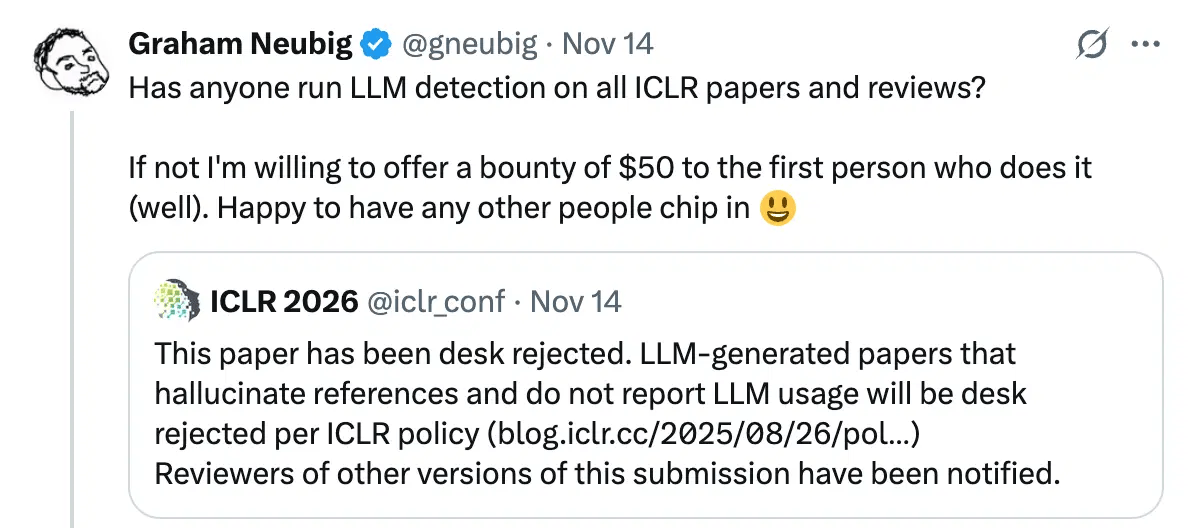}
            \captionsetup{font=small}
    \caption{Tweet on AI-generated Papers and Reviews at ICLR}
    \label{fig:ICLR}
\end{figure}

Notably close to the SE community is the recent case involving ICLR'26, one of the top machine learning conferences (see Figure \ref{fig:ICLR}). An analysis of the paper reviews submitted by PC members revealed that over 15,000 peer reviews were fully AI-generated \cite{Naddaf2025B, Pangram2025iclr}. The ICLR'26 case serves as a \textit{canary in the coal mine} for our community on the risks and potential negative consequences of the rising imbalance between paper submissions and reviewing capacity.

\subsection{Barriers to Entry}

Reviewing papers for conferences and journals constitutes a form of invisible labor with little career recognition in our “Publish or Perish” research environment. The unpaid, volunteer nature of peer review creates a fundamental asymmetry: reviewers donate substantial time and expertise to the scientific community, yet this contribution remains largely unrecognized and unrewarded within institutional career advancement systems. A recent survey of peer reviewing in software engineering by Ernst et al. \cite{Ernst2021} showed that early-career researchers prioritize reviewing for venues to determine what makes a paper strong enough to be published at top-tier venues. However, this learning opportunity comes at a personal cost, as the time investment in reviewing directly competes with their own research and publication activities.

At many institutions, reviewing is not meaningfully valued as service within tenure and promotion processes. As a result, researchers are not incentivized to serve as reviewers, as it often does not contribute directly to the advancement of their career goals \cite{Ernst2021}. This lack of institutional recognition is particularly problematic for early-career researchers (ECRs) who face pressure to maximize their publication output and secure independent funding. In such a competitive landscape, investing significant time in unrewarded service work becomes a luxury that many ECRs cannot afford. 

The existing mechanisms for recognizing reviewer contributions are insufficient to address these barriers. While valuable, Distinguished Reviewer Awards are usually awarded to more experienced members of the community, with less than 3\% of award recipients having fewer than 5 years of experience~\cite{Ernst2021}. Therefore, these awards function as a form of validation for more established researchers rather than as an incentive for attracting newcomers to the peer-reviewing process. 

Finally, existing mentorship initiatives such as the Junior/Shadow PC programs currently have important limitations that restrict their impact. The Shadow PC program, pioneered by MSR'21 and subsequently adopted by ICSE, provides valuable hands-on experience for ECRs. However, these programs are typically limited to a single review cycle and have not been widely implemented across conferences and journals, thus significantly limiting their reach and scalability. As a result, only a small subset of the growing population of ECRs can benefit from these mentorship opportunities, leaving the vast majority without structured guidance on how to conduct high-quality reviews or establish themselves as valued members of the peer review community.

\section{A More Sustainable Future of Peer Review in SE}

In this section, we provide a set of mechanisms whose adoption can build a more sustainable peer review ecosystem in the SE research community.

\subsection{Scalable Training of Junior Reviewers}

Several conferences have previously hosted presentations by experienced reviewers from the community about writing high-quality peer reviews. However, these efforts have never been broadly available to the wider community.

A more sustainable future for SE peer review has to include ways of making training for junior peer reviewers more scalable. This could be achieved by creating an online training module on writing high-quality reviews for research papers, akin to the CITI training for Responsible Conduct of Research and Human Subjects Research \cite{CITI}. The training would include videos from award-winning reviewers from the SE community and PC chairs and would systematically address the various quality aspects of a paper: motivation, novelty, methodology, presentation, etc.; reasons to reject/accept a paper; common do's and don'ts; what to do when you observe low-quality reviews; identifying AI-generated papers; responsible and allowable AI use in peer reviewing; and examples of high-quality and low-quality reviews. 

The training could also be complemented by an online platform/forum, where trainees could review some sample published papers or draft papers of their peers and receive feedback on their reviews from fellow trainees, as well as more senior members of the community. 

The online course would include a quiz at the end and issue a certificate of completion, and then post the names of the graduating trainees on a public website, along with their affiliation, research expertise, Google Scholar link, reviewing experience, and reviewing availability. In addition, reviewers could update their social media or ORCID profiles to add their certificate of completion as well as their availability, etc. This information could then be used by PC chairs and editors looking for reviewers. 

\subsection{Responsible Use of AI in the Peer Review Process}

As mentioned earlier, the use of Gen-AI by reviewers to produce fully AI-generated reviews of research papers is a growing and 
worrying issue. Although recent work on the use of agentic peer review tools has shown promising results \cite{Liang2024, Ding2025, paperreviewAI}, AI should not be used to replace the reviewers' expertise in the field or to bias their professional opinion about a paper. 

Nonetheless, we see value and potential benefits in cautiously and responsibly using Gen-AI as a tool to assist in the peer reviewing process.

First, journal editors and PC chairs should clearly define the allowed and forbidden uses of AI in paper submissions and peer reviews. As an example, the ICLR 2026 chairs allowed authors and reviewers to use AI tools to polish text, generate experiment codes, or analyze results, but required disclosure of such uses and also prohibited AI use that would breach the confidentiality of manuscripts or produce falsified content \cite{Naddaf2025B}. Next, editors and PC chairs could use AI tools such as Pangram Labs's EditLens \cite{Naddaf2025B, Thai2025} to assess the degree of AI-generated content in both paper submissions and peer reviews. Violations of the established policies should be penalized in order to discourage future incidents. This would benefit the community by reducing the number of low-quality submissions to be reviewed as well as deterring reviewers from using AI to produce their reviews.

Further, after all reviewers turn in their own, human-written reviews, AI could be used to generate an additional, complementary review to be used by reviewers to uncover potential issues that may have been overlooked. Most importantly, the AI-generated review should not be available before all the reviewers have provided their expert reviews and have had an initial discussion about the paper; this is crucial in order to avoid the introduction of bias or undue influence on the reviewers' judgments. AI could also be used to craft starting drafts for  meta-reviews, which would then be validated, modified, and expanded by the reviewers to appropriately summarize their discussion and assessment of the manuscript.

Last but not least, in order to adhere to ACM and IEEE guidelines for Gen-AI use during the review process, we would need to have community discussions with these organizations to determine the extent to which Gen-AI use is permitted and enforce the use of enterprise versions of LLMs or other third-party systems that guarantee the confidentiality of that information in order to preserve the confidential nature of the papers being reviewed \cite{ACMpeer}. 

%A valid concern would be that these restrictions could imply an additional cost for the organizing committees of SE conferences. However, this is not a concern an increasing number of universities have access to Microsoft’s enterprise level LLM (Copilot AI) as part of their Microsoft 365 license \cite{MSFT}.

\subsubsection{Incentives}
It stands to reason that more experienced researchers, who received and wrote many reviews, can provide better reviews and thus receive the Distinguished Reviewer Awards more often \cite{Ernst2021}. However, this dynamic diminishes its effectiveness in attracting new reviewers.

We believe that the following strategies to build upon and complement the existing incentives for reviewers in the community would be beneficial in the llong term:

\begin{itemize}
    \item Submission requirement: Adding a requirement for authors submitting a paper to the conference/venue to serve as reviewers as well. This would mirror the practice in NLP conferences such as ACL \cite{ACL}. Authors would have to go through the training module previously mentioned to ensure they have the required know-how to provide high-quality reviews. 
    \item Registration Discount: Providing a small discount for conference registration to members of the community serving as reviewers.
    \item Reviewer badges: Adding a visible stamp on registration badges that identify members of the program committee and Distinguished Reviewer Award recipients.
    \item Distinguished Newbie Reviewer Award: Expanding the Distinguished Reviewer Award by reserving a small fraction of the awards to be exclusively for newcomers in the community. 
\end{itemize}

\section{Conclusion}

The sustainability of peer review is a critical challenge facing the software engineering research community. In this paper, we present our vision to improve the peer review process our community relies upon and move us towards a more sustainable and rewarding peer review system that supports the continued growth and impact of the community through high-quality reviews.

Realizing this vision of the future of peer review in the SE research community will require coordinated efforts from program committee chairs, journal editors, professional organizations, and researchers in the community, but it is a viable and necessary path to ensure the survival of the peer review process in our community. 

\balance
%%
%% The next two lines define the bibliography style to be used, and
%% the bibliography file.
\bibliographystyle{ACM-Reference-Format}
\bibliography{software}

\end{document}